\documentclass{article}
\usepackage[T1]{fontenc}
\usepackage[latin9]{inputenc}
\usepackage{verbatim}
\usepackage{float}
\usepackage{amsmath}
\usepackage{amssymb}

\makeatletter

\floatstyle{ruled}
\newfloat{algorithm}{tbp}{loa}
\providecommand{\algorithmname}{Algorithm}
\floatname{algorithm}{\protect\algorithmname}

\usepackage{spconf}\usepackage{amsthm}\usepackage{breakurl}\usepackage{cite}\usepackage{amsthm}\usepackage{dsfont}\usepackage{array}\usepackage{mathrsfs}\usepackage{comment}

\title{An Algorithm for Exact Super-Resolution and Phase Retrieval}
\name{Yuxin Chen$^{\star}$  \qquad Yonina C. Eldar$^{\dagger}$ \qquad Andrea J. Goldsmith$^{\star}$ \thanks{This work is supported in part by the NSF Center for Science of Information and BSF Transformative Science Grant 2010505. }}

\address{$^{\star}$ Department of Electrical Engineering, 
Stanford University \\
    $^{\dagger}$ Department of Electrical  Engineering, 
Technion, Israel Institute of Technology}

\usepackage{hyperref}

\makeatother

\begin{document}
\ninept

\maketitle

\theoremstyle{plain}\newtheorem{lem}{\textbf{Lemma}}\newtheorem{theorem}{\textbf{Theorem}}\newtheorem{fact}{\textbf{Fact}}\newtheorem{corollary}{\textbf{Corollary}}\newtheorem{assumption}{\textbf{Assumption}}\newtheorem{example}{\textbf{Example}}\newtheorem{definition}{\textbf{Definition}}\newtheorem{proposition}{\textbf{Proposition}} 
\begin{abstract}
We explore a fundamental problem of super-resolving a signal of interest
from a few measurements of its low-pass magnitudes. We propose a 2-stage
tractable algorithm that, in the absence of noise, admits perfect
super-resolution of an $r$-sparse signal from $2r^{2}-2r+2$ low-pass
magnitude measurements. The spike locations of the signal can assume
any value over a continuous disk, without increasing the required
sample size. The proposed algorithm first employs a conventional super-resolution
algorithm (e.g. the matrix pencil approach) to recover \emph{unlabeled}
sets of signal correlation coefficients, and then applies a simple
sorting algorithm to disentangle and retrieve the true parameters
in a deterministic manner. Our approach can be adapted to multi-dimensional
spike models and random Fourier sampling by replacing its first step
with other harmonic retrieval algorithms.

\end{abstract}

\section{Introduction}

Recovering fine-grained details of an object from its coarse-scale
measurements, often termed ``super-resolution'', is a fundamental
scientific problem that arises in many signal processing problems,
e.g. direction of arrival analysis \cite{RoyKailathESPIRIT1989},
wireless channel estimation \cite{tureli2000ofdm}, medical imaging
\cite{kennedy2006super}, and optics \cite{szameit2012sparsity},
to name just a few. Due to physical limitations (e.g. diffraction
limits) and hardware constraints, it is often difficult (and sometimes
impossible) to obtain accurate measurements of the high-end spectrum
of a signal. It is thus of significant interest to super-resolve a
signal from its low-pass signal components.

Perfect super-resolution is in general impossible unless the object
of interest has a parsimonious structural representation. Many parametric
methods have been proposed to exploit the underlying harmonic structure,
including MUSIC \cite{Schmidt1986MUSIC}, ESPRIT \cite{RoyKailathESPIRIT1989},
and the matrix pencil method \cite{sarkar1995using}. These methods
are typically based on the eigen-decomposition of a matrix constructed
from low-pass samples, which can recover a signal with infi{}nite
precision in the absence of noise. Inspired by the success in sparse
recovery, Gazit et. al. \cite{gazit2009super} developed an iterative
method called Nonlocal Hard Thresholding (NLHT) for empirical super-resolution.
Candès et. al. \cite{CandesFernandez2012SR,CandesFernandez2012SRNoisy}
have recently proposed an efficient non-parametric approach based
on semidefinite programming which, under certain separation conditions,
enables provably exact and stable recovery.

The super-resolution task is more challenging in the high-frequency
regime (the regime where the carrier frequency itself is ultra-high).
Due to hardware limitations, it might be very difficult to measure
and record the phase information from low-pass magnitude measurements
in a reliable manner. This issue arises in diverse applications including
X-ray crystallography \cite{millane1990phase}, optics \cite{trebino1993using}.
and diffraction imaging \cite{szameit2012sparsity}. For many applications,
including the examples just stated, recovering the ground truth in
an efficient fashion without phase information is by nature very difficult
and oftentimes ill-posed.

Most conventional methods (e.g. the Gerchberg-Saxton algorithm \cite{gerchberg1972practical})
to recover the phase information from magnitude measurements (termed
``phase retrieval'') are based on strong prior information on the
signal, and are unable to generate provably accurate signal recovery.
A recent line of work \cite{shechtman2011sparsity,candes2013phase,candes2012phaselift}
has approached this problem from a different angle by converting the
phase retrieval problem to completion of a rank-1 lifted matrix. In
particular, Candès et. al. \cite{candes2012phaselift,candes2012solving}
deliver the encouraging performance guarantees that phase information
of any $N$-dimensional signal can be perfectly recovered from $O\left(N\right)$
random amplitude samples via efficient semidefinite programming. Stability
and uniqueness have also been studied by Eldar et. al. \cite{eldar2012phase}.
This early success has inspired a recent explosion of work in phase
retrieval \cite{li2012sparse,shechtman2013gespar,jaganathan2012recovery,alexeev2012phase,jaganathan2013sparse,chen2013PR},
from both theoretical and algorithmic perspectives.

Nevertheless, most performance guarantees for \emph{tractable }algorithms\emph{
}are established for Gaussian sampling \cite{candes2012phaselift,candes2012solving,li2012sparse,netrapalli2013phase} or
sub-Gaussian sampling \cite{chen2013PR}. Phase retrieval from Fourier intensity
measurements -- which is the most practically demanding problem --
has not been fully explored. Recent work by Jaganathan et. al. \cite{jaganathan2012recovery}
proposed a tractable algorithm that, in the presence of a \emph{full
discrete} Fourier ensemble, allows provably exact recovery for signals
up to sparsity $O\left(n^{1/3}\right)$ and empirical recovery for
signals up to sparsity $O\left(\sqrt{n}\right)$. Uniqueness has also
been investigated in \cite{ohlsson2013conditions}. However, these works do not provide provably accurate algorithms that
allow efficient recovery from the intensities of \emph{unmasked low-pass}
Fourier coefficients.

In this paper, we design an algorithm that, in the absence of noise,
allows us to retrieve the fine-scale structure of an object from the
intensities of its low-end spectrum. Our algorithm operates under
a very general setting, and enables perfect super-resolution of an
$r$-sparse signal from $m\geq2r^{2}-2r+2$ magnitude measurements
in an efficient and deterministic manner. The proposed algorithm is
a 2-step method that involves a matrix pencil approach followed by
a simple sorting algorithm. Its computational complexity is no greater
than the complexity of performing elementary inversion and eigen-decomposition
of an $\left(\frac{m}{2}-1\right)\times\left(\frac{m}{2}-1\right)$
Hankel matrix. The signal of interest can be an arbitrary \emph{continuous-time}
sparse signal, i.e. the positions of the time-domain spike of the
signal can assume any value over a continuous region.

\section{Problem Formulation: 1-D Model\label{sub:Formal-Setup}}

\subsection{Continuous-Time Model}

Assume that the continuous-time signal of interest $x(t)$ ($t\in\left[0,1\right]$)
can be modeled as a weighted superposition of spikes at $r$ distinct
positions $t_{l}$ ($1\leq l\leq r$) as follows 
\begin{equation}
x\left(t\right)=\sum_{l=1}^{r}a_{l}\delta\left(t-t_{l}\right),\label{eq:data_model}
\end{equation}
where $a_{l}$'s represent the complex amplitudes. The positions $t_{l}$
can assume any value within the continuous interval $\left[0,0.5\right)$.
The restriction is in order to avoid ambiguity, as discussed in Section
\ref{sub:Step-1:-Recovering}. Expanding $x(t)$ in a Fourier series
over the interval $[0,1]$ results in the Fourier coefficients 
\begin{equation}
\forall k\in\mathbb{Z}:\quad\hat{x}[k]=\sum_{l=1}^{r}a_{l}e^{-j2\pi kt_{l}}.\label{eq:data_model-Fourier}
\end{equation}

Suppose that we obtain magnitude information of a few frequency samples
of $x\left(t\right)$ in the low end of its spectrum, i.e. we observe
\begin{equation}
y[k]:=\left|\hat{x}[k]\right|^{2},\quad\text{ }-m_{\mathrm{c}}\leq k<m_{\mathrm{c}}.\label{eq:magnitude_model}
\end{equation}
The question is how to recover the true signal $x(t)$ from the magnitude
of these $m=2m_{\mathrm{c}}$ low-end Fourier coefficients.

\subsection{Discrete-Time Model}

The model presented in (\ref{eq:data_model}) and (\ref{eq:data_model-Fourier})
has a counterpart in the discrete-time setting as follows. Suppose
that a discrete-time signal $x[n]$ of length $N$ is a weighted sum
of $r$ spikes as follows 
\begin{equation}
x\left[n\right]=\sum_{l=1}^{r}a_{l}\delta\left[n-n_{l}\right],\label{eq:data_model-DFT}
\end{equation}
where $n_{l}\in\left\{ 0,1,\cdots,\left\lfloor \frac{N-1}{2}\right\rfloor \right\} $.
The discrete Fourier transform (DFT) coefficients of $x[n]$ is given
by 
\begin{equation}
\hat{x}\left[k\right]=\sum_{l=1}^{r}a_{l}e^{-j2\pi k\frac{n_{l}}{N}},\quad0\leq k<N.\label{eq:data_model-Fourier-DFT}
\end{equation}
Therefore, the discrete-time setting (\ref{eq:data_model-Fourier-DFT})
can be treated as a special case of the continuous-time model (\ref{eq:data_model-Fourier})
by restricting the spike positions to a fine grid $\left\{ \frac{l}{N}:0\leq l<\frac{N-1}{2N}\right\} $.
For this reason, we restrict our analysis and results to continuous-time
models, which is more general. As we will show, our method results
in recovery from the same sample complexity as for the discrete-time
model.

\section{Methodology\label{sub:Method}}

We propose a 2-step algorithm that allows perfect super-resolution
to within infinite precision, provided that the number $m$ of samples
obeys $m\geq2r^{2}-2r+2$. The proposed algorithm works under the
following fairly general conditions: 
\begin{enumerate}
\item $\left|a_{i}\right|\neq\left|a_{l}\right|$ for all $i\neq l$; 
\item $t_{i_{1}}-t_{l_{1}}\neq t_{i_{2}}-t_{l_{2}}$ for any $(i_{1},l_{1})\neq(i_{2},l_{2})$. 
\end{enumerate}
Without loss of generality, we assume that 
\begin{equation}
\left|a_{1}\right|>\left|a_{2}\right|>\cdots>\left|a_{r}\right|>0.\label{eq:a_i_order}
\end{equation}
Our algorithm first recovers \emph{unlabeled} sets of correlation
coefficients $\left\{ \left.a_{i}a_{l}^{*}\right|i\neq l\right\} $
and $\left\{ t_{i}-t_{l}\right\} $ using a matrix pencil approach,
and then retrieves the $a_{i}$'s and $t_{i}$'s via a simple sorting
algorithm. These two steps are described in more details in the following
subsections.

\subsection{Step 1: Recovering unlabeled sets of correlation coefficients via
a matrix pencil approach\label{sub:Step-1:-Recovering}}

The key observation underlying our algorithm is that 
\begin{align}
y[k]: & =\left|\hat{x}(k)\right|^{2}=\sum_{i=1}^{r}\sum_{l=1}^{r}\underset{a_{i,l}}{\underbrace{a_{i}a_{l}^{*}}}\exp\Big(-j2\pi\underset{t_{i,l}}{\underbrace{\left(t_{i}-t_{l}\right)}}k\Big).\label{eq:SquareModel}
\end{align}
Therefore, $y[k]$ corresponds to a weighted superposition of no more
than $r^{2}-r+1$ spikes in the time domain%
\footnote{Note that there are $r$ terms (i.e. all terms with $i=l$) in (\ref{eq:SquareModel})
leading to spikes at $t=0$.%
}. Recall our assumption that $t_{i}\in\left[0,0.5\right)$, which
implies that $t_{i}-t_{j}\in\left(-0.5,0.5\right)$. Since we can
only observe $y[k]$ at integer values $k\in\mathbb{Z}$, restricting
$t_{i}-t_{j}$ to within $\left(-0.5,0.5\right)$ is necessary to
avoid ambiguity.

The form (\ref{eq:SquareModel}) allows us to apply the matrix pencil
method \cite{sarkar1995using,Hua1992} to retrieve $\left\{ a_{i}a_{l}^{*}\mid i\neq l\right\} $
and $\left\{ t_{i}-t_{l}\mid i\neq l\right\} $, which we briefly
summarize as follows. Arrange $\boldsymbol{y}:=\left[y[-m_{\mathrm{c}}],\cdots,y[m_{\mathrm{c}}-1]\right]$
into an enhanced $\left(m_{\mathrm{c}}+1\right)\times m_{\mathrm{c}}$
Hankel matrix 
\begin{equation}
\small\boldsymbol{Y}:=\left[\begin{array}{cccc}
y\left[-m_{\mathrm{c}}\right] & y\left[1-m_{\mathrm{c}}\right] & \cdots & y\left[-1\right]\\
y\left[1-m_{\mathrm{c}}\right] & y\left[2-m_{\mathrm{c}}\right] & \cdots & y\left[0\right]\\
\vdots & \vdots & \ddots & \vdots\\
y\left[0\right] & y\left[1\right] & \cdots & y\left[m_{\mathrm{c}}-1\right]
\end{array}\right].\label{eq:HankelSquare}
\end{equation}
The analysis framework of \cite{sarkar1995using,Hua1992} suggests
that 
\begin{equation}
\mathrm{rank}\left(\boldsymbol{Y}\right)\leq\min\left\{ r^{2}-r+1,m_{\mathrm{c}}\right\} .\label{eq:rankY}
\end{equation}

Let $\boldsymbol{Y}_{1}$ and $\boldsymbol{Y}_{2}$ represent the
first $m_{\mathrm{c}}$ rows and the last $m_{\mathrm{c}}$ rows of
$\boldsymbol{Y}$, respectively, then both $\boldsymbol{Y}_{1}$ and
$\boldsymbol{Y}_{2}$ have rank at most $r^{2}-r+1$. The matrix pencil
method for super-resolution then proceeds as follows: 
\begin{itemize}
\item Calculate the eigenvalues $\left\{ \lambda_{i}\mid1\leq i\leq r^{2}-r+1\right\} $
of $\boldsymbol{Y}_{1}^{\dagger}\boldsymbol{Y}_{2}$, where $\boldsymbol{Y}_{1}^{\dagger}$
represents the pseudo-inverse of $\boldsymbol{Y}_{1}$; 
\item Let $\tilde{t}_{k}:=\frac{1}{2\pi}\mathrm{arg}\lambda_{k}$. One can
verify that 
\begin{equation}
\left\{ \tilde{t}_{k}\right\} =\left\{ 0\right\} \cup\left\{ \left.t_{i}-t_{l}\right|i\neq l\right\} .\label{eq:SetT}
\end{equation}
Note that we can only derive the above set in an unlabeled manner,
i.e. we are unable to link each value $\tilde{t}_{k}$ with a pair
of indices $(i,l)$ such that $\tilde{t}_{k}=t_{i}-t_{l}$. 
\item By substituting all values $\tilde{t}_{k}$ (or, equivalently, $\left\{ 0\right\} \cup\left\{ \left.t_{i}-t_{l}\right|i\neq l\right\} $)
into (\ref{eq:SquareModel}), one can derive the set of complex amplitudes
associated with each value $t_{i}-t_{l}$. In particular, one can
easily see that this amplitude set contains the values $\sum\nolimits _{i=1}^{r}\left|a_{i}\right|^{2}$
and $a_{i}a_{l}^{*}$ for all $i\neq l$. Since $\sum_{i=1}^{r}\left|a_{i}\right|^{2}$
has the largest amplitude among the whole set, we are able to separate
out the \emph{unlabeled} correlation set $\left\{ \left.a_{i}a_{l}^{*}\right|i\neq l\right\} $. 
\end{itemize}
Note that the matrix pencil approach is only one method to recover
$\left\{ \left.a_{i}a_{l}^{*}\right|i\neq l\right\} $ and $\left\{ t_{i}-t_{l}\right\} $
from a mixture of sinusoids (\ref{eq:SquareModel}). Since the spikes
at positions $t_{i}-t_{l}$ and $t_{l}-t_{i}$ always arise in pairs
in (\ref{eq:SquareModel}), we essentially only obtain $\left\{ \left|t_{i}-t_{l}\right|\right\} $.

In the absence of noise, the matrix pencil approach allows recovery
to within arbitrary precision without additional assumptions on the
spike locations. Another alternative is the total variation minimization
method recently proposed by Candès et. al. \cite{CandesFernandez2012SR,CandesFernandez2012SRNoisy},
which often improves stability in the presence of noise.

\subsection{Step 2: Recovering the $a_{i}$'s and $t_{i}$'s from the sets of
correlation coefficients via sorting}

We now consider how to retrieve $t_{i}$ and $a_{i}$ in order to
recover $x(t)$. Note that knowledge of $\left\{ \left.a_{i}a_{l}^{*}\right|i\neq l\right\} $
gives us the information on 
\begin{equation}
\left\{ \left|a_{i}a_{l}\right|:i\neq l\right\} .
\end{equation}

Suppose first that we are able to identify $\left|a_{1}\right|$.
When one knows the whole (unlabeled) set of pairwise products, various
tractable methods have been proposed to perfectly recover all $\left|a_{i}\right|$'s
(e.g. \cite{chen2009reconstructing}). Here, we employ a simple sorting
algorithm as presented in Algorithm \ref{alg:Sorting-algorithm}.
This method is based on the simple observation that the largest element
in $\left\{ \left|a_{i}a_{l}\right|:l\geq k\right\} $ is necessarily
equal to $\left|a_{1}a_{k}\right|$.

\begin{algorithm}
$\quad\quad$1. Sort $\mathcal{A}=\left\{ \left|a_{i}a_{l}\right|\mid i\neq l\right\} $.

$\quad\quad$2. \textbf{for} $2\leq i\leq r$:

$\quad\quad$3. $\quad$Set $\left|a_{i}\right|=\frac{1}{\left|a_{1}\right|}\max_{\tilde{a}\in\mathcal{A}}\tilde{a}$.

$\quad\quad$4. $\quad$\textbf{for $1\leq l<i$}:

$\quad\quad$5. $\quad\quad$$\mathcal{A}\leftarrow\mathcal{A}\backslash\left\{ \left|a_{l}a_{i}\right|\right\} $

$\quad\quad$6. $\quad$\textbf{end}

$\quad\quad$7. \textbf{end}

\caption{\label{alg:Sorting-algorithm}Sorting algorithm to recover $\left\{ \left|a_{i}\right|:1\leq i\leq r\right\} $
from $\left\{ \left|a_{i}a_{l}\right|:i\neq l\right\} $ and $\left|a_{1}\right|$.}
\end{algorithm}

Recovering $\left|a_{i}\right|$ is a crucial step since it allows
us to label the whole set $\left\{ \left|t_{i}-t_{l}\right|\mid i\neq l\right\} $.
In fact, from now on we not only have information on the set $\left\{ \left|a_{i}a_{l}\right|:i\neq l\right\} $
but also the labels (i.e. $(i,l)$) associated with all elements in
it. This immediately reveals information%
\footnote{Note that we are only able to recover the absolute value of each difference.
This arises because for any $i\neq l$, the values $t_{i}-t_{l}$
(resp. $a_{i}a_{l}^{*}$) and $t_{l}-t_{i}$ (resp. $a_{l}a_{i}^{*}$)
always come up in pairs in (\ref{eq:SquareModel}). %
} on all $\left|t_{i}-t_{l}\right|$. Recovering $t_{i}$ from all
pairwise absolute differences $\left|t_{i}-t_{l}\right|$ is now a
special case of the classical graph realization problem from Euclidean
distance \cite{dattorro2005convex,javanmard2013localization}, which
can be easily solved. Specifically, let us define $\boldsymbol{t}=[t_{1},\cdots,t_{r}]^{T}$,
an $r\times r$ distance matrix $\boldsymbol{D}$ such that 
\begin{equation}
\boldsymbol{D}_{il}=\left(t_{i}-t_{l}\right)^{2},\quad1\leq i,l\leq r,\label{eq:defn_DG}
\end{equation}
an $r\times r$ Gram matrix $\boldsymbol{G}=\boldsymbol{t}\boldsymbol{t}^{T}$,
and a geometric centering matrix $\boldsymbol{V}=\boldsymbol{I}-\frac{1}{r}{\bf 1}{\bf 1}^{T}$.
Since $\boldsymbol{D}$ is now given, one can show that (see \cite{dattorro2005convex})
\[
\boldsymbol{V}\boldsymbol{G}\boldsymbol{V}=-\boldsymbol{V}\boldsymbol{D}\boldsymbol{V}/2.
\]

The goal is to recover $\boldsymbol{G}$, which in turn allows us
to produce $\boldsymbol{t}$. However, it has been pointed out in
\cite{javanmard2013localization} that for any two Gram matrices $\boldsymbol{G}$
and $\tilde{\boldsymbol{G}}$ associated with $\left\{ t_{1},\cdots,t_{r}\right\} $
and $\left\{ \tilde{t}_{1},\cdots,\tilde{t}_{r}\right\} $ respectively,
the identity $\boldsymbol{V}\boldsymbol{G}\boldsymbol{V}=\boldsymbol{V}\tilde{\boldsymbol{G}}\boldsymbol{V}$
implies that $\left\{ t_{1},\cdots,t_{r}\right\} $ is equivalent
to $\left\{ \tilde{t}_{1},\cdots,\tilde{t}_{r}\right\} $ up to rigid
transform (i.e. rotation and translation, see \cite{javanmard2013localization}
for these definitions). Note that in our case, the rigid transform
corresponds to the global phase information that is impossible to
recover from the magnitude information.

The recovery procedure then proceeds as follows. By computing the
largest eigenvalue $\lambda_{1}$ of $\boldsymbol{V}\boldsymbol{G}\boldsymbol{V}$
and its associated eigen-vector $\boldsymbol{u}_{1}$, we obtain the
entire family of candidate solutions for $\boldsymbol{t}=[t_{1},\cdots,t_{l}]^{T}$
that yield the same $\boldsymbol{V}\boldsymbol{G}\boldsymbol{V}$
as follows 
\begin{equation}
\tilde{\boldsymbol{t}}=\sqrt{\lambda_{1}}\boldsymbol{u}_{1}+c_{1}\boldsymbol{1},\quad\text{or}\quad\tilde{\boldsymbol{t}}=-\sqrt{\lambda_{1}}\boldsymbol{u}_{1}+c_{2}\boldsymbol{1}.\label{eq:CandidateSolutionT}
\end{equation}
Here, $c_{1},c_{2}$ are arbitrary scalars that encode the global
shift of spike positions. Note that all of these candidates satisfying
$\tilde{\boldsymbol{t}}\in\left[0,0.5\right)^{r}$ are valid solutions
compatible with the measurements. In practice, the solutions can be
refined with the aid of information on a few (two or more) reference
/ anchor spikes.

After we retrieve the $t_{i}$'s, we equivalently derive all labels
for $\mathrm{arg}\left(a_{i}a_{l}^{*}\right)=\mathrm{arg}\left(a_{i}\right)-\mathrm{arg}\left(a_{l}\right)$.
Since we have knowledge on all values of $\mathrm{arg}\left(a_{i}\right)-\mathrm{arg}\left(a_{l}\right)$,
recovering $\mathrm{arg}\left(a_{i}\right)$ can then be easily solved
by elementary linear algebra, except for a global phase on the $a_{i}$'s.

\begin{algorithm*}
1. Using the matrix pencil approach to retrieve the sets $\left\{ \left.a_{i}a_{l}^{*}\right|i\neq l\right\} $
and $\left\{ t_{i}-t_{l}\mid i\neq l\right\} $.

$\quad\quad$(a) Calculate the eigenvalues $\left\{ \lambda_{i}\right\} $
of $\boldsymbol{Y}_{1}^{\dagger}\boldsymbol{Y}_{2}$, where $\boldsymbol{Y}_{1}$
and $\boldsymbol{Y}_{2}$ are the first and the last $m_{\mathrm{c}}$
rows of $\boldsymbol{Y}$ of (\ref{eq:HankelSquare}), respectively.

$\quad\quad$(b) Let $\tilde{t}_{i}:=\frac{1}{2\pi}\mathrm{arg}\lambda_{i}$.
Then the set $\left\{ t_{i}-t_{l}\right\} =\left\{ \tilde{t}_{i}\right\} $.

$\quad\quad$(c) Substitute $\left\{ t_{i}-t_{l}\right\} $ into (\ref{eq:SquareModel})
to obtain $\left\{ \left.a_{i}a_{l}^{*}\right|i\neq l\right\} $.

\vspace{5pt}

2. \textbf{Initialize} $\mathcal{S}=\left\{ \left|a_{i}a_{l}\right|:i\neq l\right\} $,
and set $\left|a_{1}a_{2}\right|$ and $\left|a_{1}a_{3}\right|$
to be the largest 2 elements of $S$. $\mathcal{S}\leftarrow\mathcal{S}\backslash\left\{ \left|a_{1}a_{2}\right|,\left|a_{1}a_{3}\right|\right\} $.

\textbf{$\quad\quad$for $i=1:r-2$}

$\quad\quad\quad$let $s^{*}=\max_{s\in\mathcal{S}}\mathcal{S}$,
and let $\mathcal{S}\leftarrow\mathcal{S}\backslash\left\{ s^{*}\right\} $

$\quad\quad\quad$set $\left|a_{1}\right|\leftarrow\sqrt{\left|a_{1}a_{2}\right|\left|a_{1}a_{3}\right|}/\sqrt{s^{*}}$.

$\quad\quad\quad$Perform Algorithm \ref{alg:Sorting-algorithm} to
identify $\left\{ \left|a_{i}\right|\right\} $, which in turn allows
us to retrieve $\left|t_{i}-t_{l}\right|$ for all $i\neq l$.

$\quad\quad\quad$Let $\boldsymbol{D}:=[(t_{i}-t_{l})^{2}]_{1\leq i,l\leq r}$
and $\boldsymbol{V}:=\boldsymbol{I}-\frac{1}{r}{\bf 1}{\bf 1}^{T}$.
Set $\boldsymbol{G}_{\boldsymbol{V}}:=-\boldsymbol{V}\boldsymbol{D}\boldsymbol{V}/2$,
and compute its largest eigenvalue $\lambda_{1}$ and the associated
eigenvector $\boldsymbol{u}_{1}$.

$\quad\quad\quad$Obtain candidate solutions for $\boldsymbol{t}:=\{t_{1},\cdots,t_{r}\}$
are given by $\boldsymbol{t}=\sqrt{\lambda_{1}}\boldsymbol{u}_{1}+c_{1}{\bf 1}$
or $\boldsymbol{t}=-\sqrt{\lambda_{1}}\boldsymbol{u}_{1}+c_{2}{\bf 1}$
for any $c_{1}$ and $c_{2}$ that encode the global shift.

$\quad\quad\quad$Compute $\arg a_{i}$ (up to a global phase) using
all values $\arg a_{i}-\arg a_{l}$.

$\quad\quad\quad$\textbf{if }this iteration yields a valid solution
(i.e. obeying $\boldsymbol{t}\in\left[0,0.5\right)^{r}$ and (\ref{eq:magnitude_model})):

$\quad\quad\quad\quad$report this solution;

$\quad\quad\quad$\textbf{end}

\textbf{$\quad\quad$end}

\caption{Super-Resolution and Phase Retrieval Algorithm\textbf{\label{alg:Super-resolution-Algorithm}}}
\end{algorithm*}

It remains to determine $\left|a_{1}\right|$. Observe that 
\[
\forall(i,l)\notin\left\{ (1,2),(1,3)\right\} :\quad\left|a_{1}a_{2}\right|\geq\left|a_{1}a_{3}\right|\geq\left|a_{i}a_{l}\right|
\]
and 
\[
\forall i\geq2,l\geq3,\quad\left|a_{2}a_{3}\right|\geq\left|a_{i}a_{l}\right|.
\]
Therefore, $\left|a_{2}a_{3}\right|$ can only take place within the
largest $r-2$ elements of the set $\left\{ \left|a_{i}a_{l}\right|:i\neq l\right\} $
(i.e. $\left\{ \left|a_{1}a_{l}\right|\mid4\leq l\leq r\right\} \cup\left\{ \left|a_{2}a_{3}\right|\right\} $).
For each value of $\left|a_{2}a_{3}\right|$, we can easily determine
$\left|a_{1}\right|$ as follows 
\[
\left|a_{1}\right|=\sqrt{\left|a_{1}a_{2}\right|\left|a_{1}a_{3}\right|}/\left|a_{2}a_{3}\right|.
\]
An exhaustive search over all $r-2$ choices and checking compatibility
for each choice allow us to solve the problem exactly.

In summary, our algorithm is able to return all solutions compatible
with the measurements. In the cases where uniqueness is not guaranteed,
our algorithm can discover all possible solutions.

\subsection{Discussion}

\textbf{Complexity}. The proposed solution is summarized in Algorithm
\ref{alg:Super-resolution-Algorithm}. One can see that the bottleneck
lies in the matrix pencil approach, which involves inversion and eigen-decomposition
of an $m_{\mathrm{c}}\times m_{\mathrm{c}}$ matrix. Therefore, our
algorithm has computational complexity no greater than elementary
inversion and eigen-decomposition of a Hankel matrix, and it is capable
of recovering \emph{all} signals that are compatible with the magnitude
samples. That said, we do not need the uniqueness condition (e.g.
\cite{ohlsson2013conditions,ranieri2013phase}) in order to perform
recovery. The algorithm works as soon as the number $m$ of measurements
exceeds $2r^{2}-2r+2$. In other words, our algorithm admits perfect
super-resolution up to sparsity $O\left(\sqrt{m}\right)$.

\textbf{Comparison with }\cite{jaganathan2012recovery}. When a full
$N$-dimensional discrete Fourier ensemble is present, the algorithm
proposed in \cite{jaganathan2012recovery} can provably work for signals
up to sparsity $O(\sqrt{N})$, and numerically work for signals up
to sparsity $O(N^{1/3})$. The complexity of the algorithms therein
is a polynomial function of the size of the grid in which the discrete-time
signal lies, and the recovery guarantee can only be stated in a probabilistic
sense. In contrast, our algorithm can recover any continuous-time
spike with infinite precision \emph{deterministically}, and the computational
complexity depends only on the signal sparsity $r$.

\section{Extensions\label{sub:Extension}}

\subsection{Multi-dimensional Spikes}

Our method immediately extends to multi-dimensional spike models.
Suppose that $x\left(\boldsymbol{t}\right)$ is a mixture of $K$-dimensional
spikes at $r$ distinct locations $\boldsymbol{t}_{i}\in\left[0,0.5\right)^{K}$
($1\leq i\leq r$). If we let $\hat{x}[\boldsymbol{k}]$ denote the
$K$-dimensional Fourier series coefficients of $x\left(\boldsymbol{t}\right)$,
then we can write 
\[
y[\boldsymbol{k}]:=\left|\hat{x}(\boldsymbol{k})\right|^{2}=\sum\nolimits _{i,l=1}^{r}a_{i}a_{l}^{*}\exp\left(-j2\pi\left\langle \boldsymbol{t}_{i}-\boldsymbol{t}_{l},\boldsymbol{k}\right\rangle \right).
\]
Recovering the unlabeled sets $\left\{ a_{i}a_{l}^{*}\mid i\neq l\right\} $
and $\left\{ \boldsymbol{t}_{i}-\boldsymbol{t}_{l}\mid i\neq l\right\} $
can be done by multi-dimensional matrix pencil methods (e.g. \cite{Hua1992}).
Note that the matrix pencil form for $K$-dimensional spike models
is no longer a Hankel matrix, but instead an enhanced $K$-fold Hankel
matrix, as discussed in \cite{chen2013robust}.

After we identify $\left\{ a_{i}a_{l}^{*}\mid i\neq l\right\} $ and
$\left\{ \boldsymbol{t}_{i}-\boldsymbol{t}_{l}\mid i\neq l\right\} $,
then $\boldsymbol{t}_{i}$ can be retrieved in a coordinate-wise manner,
i.e. we apply the second step of Algorithm \ref{alg:Super-resolution-Algorithm}
for each coordinate and retrieve it. This generates all signals compatible
with the measurements.

\subsection{Random Fourier Sampling}

Our algorithms can also be adapted to accommodate random Fourier magnitude
samples, by replacing Step 1 with more appropriate harmonic retrieval
algorithms. For example, when the underlying spikes lie on a fine
grid, one can attempt recovery via a compressed sensing algorithm
(e.g. $\ell_{1}$ minimization in \cite{CandRomTao06}), MUSIC, or
NLHT \cite{gazit2009super}. When the spike locations can assume any
value over a continuous region, more complicated convex optimization
methods are needed to address the basis mismatch issue \cite{Chi2011sensitivity}.
Examples include the atomic norm minimization \cite{TangBhaskarShahRecht2012}
for the 1-D model and Hankel matrix completion \cite{chen2013robust}
for multi-dimensional models.

More broadly, Step 2 of Algorithm \ref{alg:Super-resolution-Algorithm}
is quite general and can build on top of any method that can retrieve
the sets $\left\{ a_{i}a_{l}^{*}\mid i\neq l\right\} $ and $\left\{ \boldsymbol{t}_{i}-\boldsymbol{t}_{l}\mid i\neq l\right\} $
from the obtained measurements, regardless of the pattern of the obtained
intensity measurements.

\section{Numerical Example\label{sub:NumericalExperiment}}

We conduct the following numerical example to illustrate the correctness
of our algorithm. Generate a signal $x(t)$ of $r=5$ random spikes
lying in $\left(0,0.5\right)$. The amplitudes associated with the
spikes are independently drawn from $\mathcal{N}\left(0,1\right)$.
Suppose we observe 
\[
y[k]:=\left|\hat{x}[k]\right|^{2}\quad(-m_{\mathrm{c}}\leq k<m_{\mathrm{c}})
\]
for various choices of $m_{\mathrm{c}}$. To avoid numerical issues,
the spike positions $t_{i}$'s are generated such that 
\begin{equation}
\min\left\{ \left|\alpha-\beta\right|\left|\alpha,\beta\in\mathcal{T}_{\text{diff}}\right.\right\} \geq0.02,\label{eq:SeparationCondition}
\end{equation}
where $\mathcal{T}_{\text{diff}}:=\left\{ \left|t_{i}-t_{l}\right|:1\leq i,l\leq r\right\} $.
Condition (\ref{eq:SeparationCondition}) is some separation condition
typically required to ensure numerical stability. In fact, we observe
that if the separation condition is violated, then the matrix pencil
approach is often numerically unstable as well.

Under the above model, the algorithm works perfectly in recovering
the underlying frequencies whenever $m\geq2r^{2}-2r+2$. For example,
when the spikes are defined by 
\[
\boldsymbol{t}=\left[0.0092,0.1411,0.3435,0.3735,0.4463\right]
\]
\[
\text{and}\quad\boldsymbol{a}=\left[0.4296,0.5160,0.9052,-0.0785,-2.2056\right],
\]
the recovery on both $\boldsymbol{a}$ and $\boldsymbol{t}$ is exact
(with inaccuracy $1.2385\times10^{-8}$) except for the global phase,
whenever $m\geq42$.

\section{Conclusion and Future Work}

We present an efficient 2-stage algorithm that allows us to super-resolve
a signal from a few Fourier intensity measurements in its low-end
spectrum. We demonstrate that for almost all signals with sparsity
$r$, the algorithm admits perfect signal recovery from as few as
$2r^{2}-2r+2$ magnitude samples. The signal spikes are not required
to lie on a fine grid, and the algorithm can be extended to accommodate
multi-dimensional spike models and random Fourier samples.

It remains to be seen whether efficient algorithms can be found to
accurately recover a sparse signal from even fewer magnitude samples.
In addition, the success of the proposed super-resolution algorithm
highly relies on the sorting algorithm, which is not very stable in
the presence of noise. An algorithm more robust to noise might need
to retrieve $t_{i}$ and $a_{i}$ simultaneously to improve stability.
It would also be interesting to explore whether there is a non-parametric
method for Step 2, i.e. to (approximately) retrieve $t_{i}$'s and
$a_{i}$'s from the unlabeled correlation sets without prior information
on the model order.

 \bibliographystyle{IEEEbib}
\bibliography{bibfileToeplitzPR}

\begin{thebibliography}{10}

\bibitem{RoyKailathESPIRIT1989}
R.~Roy and T.~Kailath,
\newblock ``{ESPRIT}-estimation of signal parameters via rotational invariance
  techniques,''
\newblock {\em IEEE Trans on Acoustics, Speech and Signal Proc.}, vol. 37, no.
  7, 1989.

\bibitem{tureli2000ofdm}
U.~Tureli, H.~Liu, and M.~D. Zoltowski,
\newblock ``{OFDM} blind carrier offset estimation: {ESPRIT},''
\newblock {\em IEEE Transactions on Communications}, vol. 48, no. 9, pp.
  1459--1461, 2000.

\bibitem{kennedy2006super}
J.~A. Kennedy, O.~Israel, A.~Frenkel, R.~Bar-Shalom, and H.~Azhari,
\newblock ``Super-resolution in {PET} imaging,''
\newblock {\em IEEE Trans on Medical Imaging}, vol. 25, no. 2, pp. 137--147,
  2006.

\bibitem{szameit2012sparsity}
A.~Szameit, Y.~Shechtman, E.~Osherovich, E.~Bullkich, P.~Sidorenko, H.~Dana,
  S.~Steiner, E.B. Kley, S.~Gazit, T.~Cohen-Hyams, S.~Shoham, M.~Zibulevsky,
  I.~Yavneh, Y.~C. Eldar, O.~Cohen, and M.~Segev,
\newblock ``Sparsity-based single-shot subwavelength coherent diffractive
  imaging,''
\newblock {\em Nature materials}, vol. 11, no. 5, pp. 455--459, 2012.

\bibitem{Schmidt1986MUSIC}
R.~Schmidt,
\newblock ``Multiple emitter location and signal parameter estimation,''
\newblock {\em IEEE Transactions on Antennas and Propagation}, vol. 34, no. 3,
  pp. 276--280, 1986.

\bibitem{sarkar1995using}
T.~K. Sarkar and O.~Pereira,
\newblock ``Using the matrix pencil method to estimate the parameters of a sum
  of complex exponentials,''
\newblock {\em IEEE Antennas and Propagation Magazine}, vol. 37, no. 1, pp.
  48--55, 1995.

\bibitem{gazit2009super}
S.~Gazit, A.~Szameit, Y.~C. Eldar, and M.~Segev,
\newblock ``Super-resolution and reconstruction of sparse sub-wavelength
  images,''
\newblock {\em Optics Express}, vol. 17, no. 26, pp. 23920--23946, 2009.

\bibitem{CandesFernandez2012SR}
E.~J. Candes and C.~Fernandez-Granda,
\newblock ``Towards a mathematical theory of super-resolution,''
\newblock {\em to appear in Communications on Pure and Applied Mathematics},
  2013.

\bibitem{CandesFernandez2012SRNoisy}
E.~J. Candes and C.~Fernandez-Granda,
\newblock ``Super-resolution from noisy data,''
\newblock November 2012.

\bibitem{millane1990phase}
R.P. Millane,
\newblock ``Phase retrieval in crystallography and optics,''
\newblock {\em JOSA A}, vol. 7, no. 3, pp. 394--411, 1990.

\bibitem{trebino1993using}
R.~Trebino and D.~J. Kane,
\newblock ``Using phase retrieval to measure the intensity and phase of
  ultrashort pulses: frequency-resolved optical gating,''
\newblock {\em JOSA A}, vol. 10, pp. 1101--1111, 1993.

\bibitem{gerchberg1972practical}
RW~Gerchberg and W.~O. Saxton,
\newblock ``A practical algorithm for the determination of phase from image and
  diffraction plane pictures,''
\newblock {\em Optik}, vol. 35, pp. 237, 1972.

\bibitem{shechtman2011sparsity}
Y.~Shechtman, Y.~C. Eldar, A.~Szameit, and M.~Segev,
\newblock ``Sparsity based sub-wavelength imaging with partially incoherent
  light via quadratic compressed sensing,''
\newblock {\em Optics Express}, 2011.

\bibitem{candes2013phase}
E.~J. Candes, Y.~C. Eldar, T.~Strohmer, and V.~Voroninski,
\newblock ``Phase retrieval via matrix completion,''
\newblock {\em SIAM Journal on Imaging Sciences}, vol. 6, no. 1, pp. 199--225,
  2013.

\bibitem{candes2012phaselift}
E.~J. Candes, T.~Strohmer, and V.~Voroninski,
\newblock ``Phaselift: Exact and stable signal recovery from magnitude
  measurements via convex programming,''
\newblock {\em Communications on Pure and Applied Mathematics}, 2012.

\bibitem{candes2012solving}
E.~J. Candes and X.~Li,
\newblock ``Solving quadratic equations via {PhaseLift} when there are about as
  many equations as unknowns,''
\newblock {\em Foundations of Computational Math}, 2013.

\bibitem{eldar2012phase}
Y.~C. Eldar and S.~Mendelson,
\newblock ``Phase retrieval: Stability and recovery guarantees,''
\newblock {\em Applied and Computational Harmonic Analysis}, September 2013.

\bibitem{li2012sparse}
X.~Li and V.~Voroninski,
\newblock ``Sparse signal recovery from quadratic measurements via convex
  programming,''
\newblock {\em SIAM Journal on Mathematical Analysis}, 2013.

\bibitem{shechtman2013gespar}
Y.~Shechtman, A.~Beck, and Y.~C. Eldar,
\newblock ``{GESPAR}: Efficient phase retrieval of sparse signals,''
\newblock {\em arXiv:1301.1018}.

\bibitem{jaganathan2012recovery}
K.~Jaganathan, S.~Oymak, and B.~Hassibi,
\newblock ``Recovery of sparse 1-{D} signals from the magnitudes of their
  {F}ourier transform,''
\newblock in {\em IEEE ISIT}, 2012, pp. 1473--1477.

\bibitem{alexeev2012phase}
B.~Alexeev, A.~S. Bandeira, M.~Fickus, and D.~G. Mixon,
\newblock ``Phase retrieval with polarization,''
\newblock {\em arXiv:1210.7752}.

\bibitem{jaganathan2013sparse}
K.~Jaganathan, S.~Oymak, and B.~Hassibi,
\newblock ``Sparse phase retrieval: Convex algorithms and limitations,''
\newblock {\em IEEE International Symposium on Information Theory}, 2013.

\bibitem{chen2013PR}
Y.~Chen, Y.~Chi, and A.~J. Goldsmith,
\newblock ``Exact and stable covariance estimation from quadratic sampling via
  convex programming,''
\newblock {\em http://arxiv.org/abs/1310.0807}, 2013.

\bibitem{netrapalli2013phase}
P.~Netrapalli, P.~Jain, and S.~Sanghavi,
\newblock ``Phase retrieval using alternating minimization,''
\newblock {\em Advances in Neural Information Processing Systems (NIPS)}, 2013.

\bibitem{ohlsson2013conditions}
H.~Ohlsson and Y.~C. Eldar,
\newblock ``On conditions for uniqueness in sparse phase retrieval,''
\newblock {\em arXiv:1308.5447}, 2013.

\bibitem{Hua1992}
Y.~Hua,
\newblock ``Estimating two-dimensional frequencies by matrix enhancement and
  matrix pencil,''
\newblock {\em IEEE Transactions on Signal Processing}, vol. 40, no. 9, pp.
  2267 --2280, Sep 1992.

\bibitem{chen2009reconstructing}
S.~Chen, Z.~Huang, and S.~Kannan,
\newblock ``Reconstructing numbers from pairwise function values,''
\newblock in {\em Algorithms and Computation}, pp. 142--152. Springer, 2009.

\bibitem{dattorro2005convex}
J.~Dattorro,
\newblock {\em Convex optimization and {E}uclidean distance geometry},
\newblock Meboo Publishing USA, 2005.

\bibitem{javanmard2013localization}
A.~Javanmard and A.~Montanari,
\newblock ``Localization from incomplete noisy distance measurements,''
\newblock {\em Foundations of Computational Math}, vol. 13, pp. 297--345, June
  2013.

\bibitem{ranieri2013phase}
J.~Ranieri, A.~Chebira, Y.~M. Lu, and M.~Vetterli,
\newblock ``Phase retrieval for sparse signals: Uniqueness conditions,''
\newblock {\em arXiv preprint arXiv:1308.3058}, 2013.

\bibitem{chen2013robust}
Y.~Chen and Y.~Chi,
\newblock ``Robust spectral compressed sensing via structured matrix
  completion,''
\newblock {\em submitted to IEEE Transactions on Information Theory}, April
  2013.

\bibitem{CandRomTao06}
E.~J. Candes, J.~Romberg, and T.~Tao,
\newblock ``Robust uncertainty principles: exact signal reconstruction from
  highly incomplete frequency information,''
\newblock {\em IEEE Transactions on Information Theory}, vol. 52, no. 2, pp.
  489--509, Feb. 2006.

\bibitem{Chi2011sensitivity}
Y.~Chi, L.L. Scharf, A.~Pezeshki, and A.R. Calderbank,
\newblock ``Sensitivity to basis mismatch in compressed sensing,''
\newblock {\em IEEE Trans on Signal Proc.}, vol. 59, no. 5, pp. 2182--2195,
  2011.

\bibitem{TangBhaskarShahRecht2012}
G.~Tang, B.~N. Bhaskar, P.~Shah, and B.~Recht,
\newblock ``Compressed sensing off the grid,''
\newblock July 2012.

\end{thebibliography}

\end{document}